# Self coupling of the Higgs boson to the Higgs field and predictions for the Higgs mass and production cross section at the Fermilab Tevatron collider

N.D. Giokaris

*Physics Department, The University of Athens, 157 71 Athens, Greece*

**Abstract**

The mass-generation mechanism is one of the most important problems in modern particle physics. The discovery and study of the Higgs boson would greatly contribute to the understanding and solving of this problem. One of the unknowns in the Higgs potential is the Higgs quadric self-coupling parameter λ. A λ parameter value equal to 1 leads to the prediction that the Standard Model Higgs mass is twice the top quark mass or about 347 GeV. It is, then, argued that this could have a dramatic increase of the SM Higgs production cross section at the Tevatron and the LHC, thus making possible its discovery even at the Tevatron with about 5 fb(-1) of integrated luminosity.

## I. Introduction

The Higgs boson and the top quark are fundamental ingredients of the Standard Model (SM) of the elementary particles and their interactions. Although the SM requires their existence it does not predict their masses in a direct way. In Ref. [1] simple predictions of the masses of these two particles were given under the assumption that both, the Yukawa coupling of the top quark to the Higgs field, and the quadric "self coupling" term λ in the Higgs potential are equal to 1. That the top quark Yukawa coupling is consistent with 1 is, by now, a well established experimental fact based on the very accurate top quark mass measurement at the Tevatron [2, 3]. In Ref. [1] it was explained how a λ parameter value of 1 leads to the SM Higgs mass being equal to two times the top quark mass or about 347 GeV. In section II an argument and a mechanism are presented according to which the fact that the SM Higgs boson mass is twice the top quark mass could lead to a dramatic increase to its production cross section at the high energy colliders, thus making possible its discovery even at the Fermilab Tevatron by the end of RUN II.

## II. Higgs boson production cross section enhancement if its mass is $m_H=2m_t$

The Standard Model Higgs particle is supposed to be produced at the Tevatron, or the LHC, mainly with gluon-gluon fusion through a t-tbar loop as shown in Fig. 1. The cross section for this process is estimated to be about ~ 0.01 pb = 10 fb at the Tevatron for a Higgs mass of ~ 340 GeV. A production cross section of this size would make it impossible to discover the Higgs by CDF and/or D0 as the total integrated luminosity, expected to be accumulated by the end of RUN II, is about 10 fb(-1) per experiment.

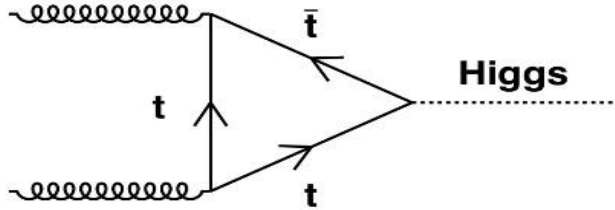

Fig. 1   Main diagram for SM Higgs production at the high energy colliders

In this paper a possible way is proposed by which the Higgs particle production cross section could be considerably enhanced (~ x100) thus making it possible to observe it at the Tevatron. The idea is based on the possibility, put forward by the authors of the paper in Ref. [1], that the Higgs boson mass is twice the top quark mass and taking into account the similarities of the Higgs production (Fig. 1) and the t-tbar main production diagram at the Tevatron as shown in Fig. 2.

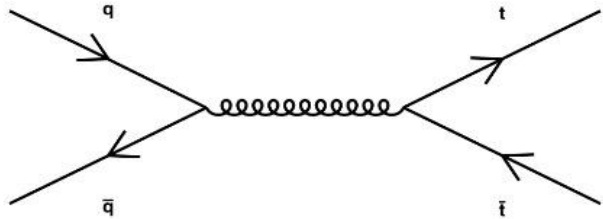

Fig. 2   Main diagram for t-tbat production at the Tevatron

Because of the strong coupling (the top quark Yukawa coupling is equal to 1) of the t-tbar pair to the Higgs field/particle it is conceivable that a significant fraction F of the t-tbars, produced according to the diagram shown in Fig. 2, come together, through the emission of a semisoft gluon g' (this gluon is also needed to neutralize the color of the t-tbar system - the original gluon carries color whereas the Higgs, the final product, is colorless) to produce a Higgs particle through a loop diagram as shown in Fig. 3. Because the emitted soft gluon g' takes away some energy (~ 5 – 10 GeV) from the t-tbar system the produced Higgs will be slightly virtual and the average value of its mass should be somewhat smaller than 2x $m_t$, i.e. ~ 335 - 340 GeV. Actually the jet from this soft gluon could help to tag this kind of events. The best place to look for this soft jet will be in ZZ events where one Z decays to two leptons and the other Z to neutrino-antineutrino without any extra jet activity. We note the close similarity of the diagram shown in Fig. 3 with that of Fig. 2. A rough estimate of the fraction F could be obtained by the formula:

$$F = \sigma(ttbar) \times (g_t \times \alpha_s)^2 \qquad (1)$$

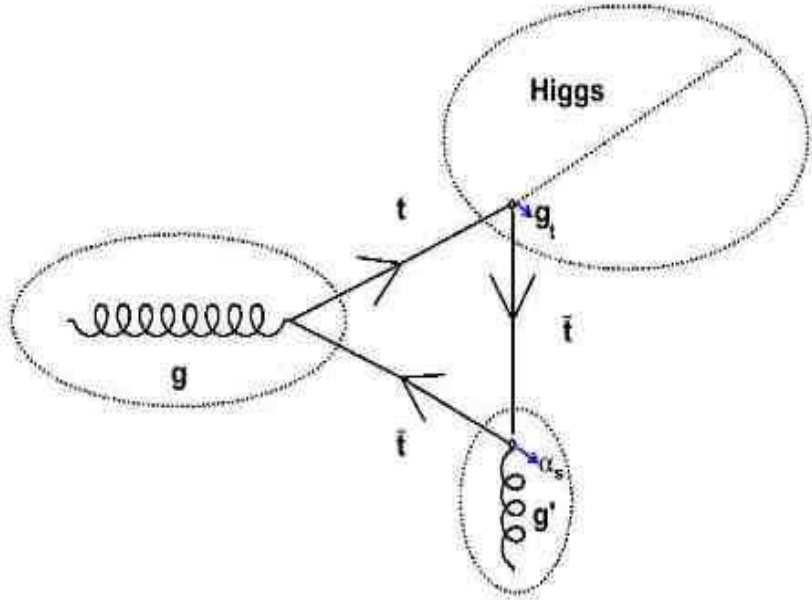

Fig.3 Higgs production through direct t-tbar pair production at the Tevatron

But $g_t = 1$ (top quark Yakawa coupling) and **$α_s$** should be large because the gluon's g' $q^2$ is rather small. If we take $α_s \sim 1/3$, then, from Eqn. 1, it is:

$$F = 7.6 \text{ pb} \times 1/9 = 0.8 \text{ pb}$$

The fact that the emitted gluon should have the right color will reduce F but this could be compensated by the fact that the gluon can be emitted from two vertices and also the value of $α_s$ can be larger. As the t-tbar pair production cross section at the Tevatron is $\sim 7.6$ pb [4] the SM Higgs could be produced at the Tevatron by this process with a cross section of about 1 pb. This could make the SM Higgs particle observable at the Tevatron even with an integrated luminosity of $\sim 5$ fb(-1). Similar boosts in the SM Higgs cross section should, of course, be in effect also at the LHC making its observation there much easier too. However it should be reminded that there is a fundamental difference between the Tevatron and the LHC machines. At Tevatron protons collide with antiprotons (matter − antimatter) whereas at the LHC protons collide with protons (matter-matter). It is not inconceivable that this could have dramatic consequences on the production of such a fundamental particle as the Higgs. The experimental data will show.

**IV.    Conclusions**

The Standard Model Higgs particle is supposed to be produced at the Tevatron mainly with gluon-gluon fusion through a t-tbar loop. The cross section for this process is about 0.01 pb for a Higgs mass of $\sim 345$ GeV. In this paper it is shown how an enhancement by a factor of about 100 of this figure could be achieved, if the Higgs mass is twice the top quark mass. An enhancement of this magnitude could make the discovery of the Higgs boson feasible at the Tevatron with about 5 fb(-1) of integrated luminosity through the H -> ZZ(star) decay channel.

**References**

(1)     "On the Higgs Mass Generation Mechanism in the Standard Model", V.A. Bednyakov, N.D. Giokaris and V.A. Bednyakov, arXiv:hep-ph/0703280v1, 27 March 2007 and published in: Physics of Particles and Nuclei, 2008, Vol. 39, No. 1, pp 13 – 36

(2)     T. Aaltonen et al., (CDF Collaboration), Phys. Rev. D 83, 111101(2011)

(3)     V. M. Abazov et al., (D0 Collaboration), Phys. Rev. D 84, 032004 (2011)

(4)     T. Aaltonen et al., (CDF Collaboration), Phys. Rev. D 84, 031101(2011)